\documentstyle[12pt,aasms4]{article}
\lefthead{Rothschild et al.}
\righthead{RXTE Observation of Cen A}
\received{13 May 1998}
\accepted{26 July 1998}
\begin{document}

\title{Observation of Cen A by the Rossi X-ray Timing Explorer}

\author{R. E. Rothschild, D. L. Band, P. R. Blanco, D. E. Gruber, W. A. Heindl, D. R. MacDonald, D. C. Marsden}
\affil{Center for Astrophysics and Space Sciences 0424, \\
University of California at San Diego, La Jolla, CA 92093}

\author{K. Jahoda, D. Pierce, G. Madejski}
\affil{Goddard Space Flight Center, Greenbelt, MD 20771}

\author{M. Elvis, D. A. Schwartz}
\affil{Center for Astrophysics,\\
Harvard-Smithsonian Astrophysical Observatory, Cambridge, MA 02138}

\author{R. Remillard}
\affil{Center for Space Research, \\
Massachusetts Institute of Technology, Cambridge, MA 02139}

\author{A. A. Zdziarski}
\affil{Nicolaus Copernicus Astronomical Center, Warsaw, Poland}

\author{C. Done}
\affil{Dept. of Physics,\\
University of Durham, Durham, England}

\and

\author{R. Svensson}
\affil{Stockholm Observatory, Saltsj\"obaden, Sweden}

\begin{abstract}
The {\it Rossi X-ray Timing Explorer} made a short (10 ks) observation
of the radio galaxy Centaurus A on 14 August 1996. Analysis of the
combined 2.5$-$240 keV spectrum  has revealed a heavily absorbed
(N$_H$=9.42$\pm$0.24$\times$10$^{22}$ cm$^{-2}$) primary power law
($\Gamma$=1.86$\pm$0.015) and an iron line due to fluorescence of cold
matter (E.W.=162$\pm$25 eV).  Flux from either a jet, primary flux
scattered into the line of sight, or primary flux seen through a
partial absorber was not required.  The iron line width is unresolved
at the 95\% confidence level ($\sigma <$ 0.54 keV).  No significant
variability in the iron line flux is seen from measurements over the
last two decades, while the overall continuum flux varied by more than
a factor of four, which implies that the line emission region is distant from
that of the primary emission. While radio-quiet Seyfert galaxies
exhibit spectral components attributable to Compton reflection from
cold matter, Cen A reveals no such component (exposed solid angle ratio
$\Omega/2\pi \leq$0.09). This supports unified models of active galaxies
that have little difference between Seyfert 2 and low luminosity radio 
galaxies.
\end{abstract}


\section{Introduction}

Cen A (NGC 5128) was discovered in X-rays over 25 years ago (Bowyer et
al.\markcite{bowyer} 1970) and has been observed by most X-ray
astronomy missions since then (see compilations in Baity et
al.\markcite{baity} 1981 and Jourdain et al.\markcite{jourdain} 1993).
At a distance of $\sim$3.5 Mpc (Hui et al.\markcite{hui} 1993), Cen A
is one of the nearest active galactic nuclei (AGN). Even though the
nucleus is heavily obscured (in part by the famous dust-lane that is
seen to cross the galaxy), Cen A is one of the brightest extragalactic
X-ray sources.  Furthermore, Cen A is clearly a radio-loud object,
displaying prominent radio jets and lobes (Burns, Feigelson, \&
Schreier\markcite{burns} 1983),  and, as such, has provided much of our
information on the high energy nature of low luminosity radio-loud
AGN.  A comparison of Cen A to the well-studied Seyfert galaxies
(Nandra \& Pounds\markcite{nandra} 1994, Nandra et
al.\markcite{nandraetal} 1997, Turner et al.\markcite{janeb,janec}
1997b,c), as well as other radio-loud AGN, in the context of the
unified model for AGN (e.g. Antonucci\markcite{antonucci} 1993,
Wo\'zniak et al.\markcite{woz} 1998), can yield insight into the
differences between radio-loud and radio-quiet AGN.

One area of radio-loud/radio-quiet difference is in the amount of the
observed Compton ``reflection''component. Compton reflection and
line fluorescence (especially from iron) are important diagnostics of
the central engine structure. Many Seyfert (i.e., radio-quiet AGN)
nuclei show these strong features which are widely believed to result
from irradiation of a thin accretion disk by a
disk corona (Guilbert \& Rees\markcite{gandr} 1988; Lightman \&
White\markcite{landw} 1988; George \& Fabian\markcite{gandf} 1991).
This component manifests itself as an iron line at 6.4 keV, an iron
edge at 7.1 keV, and a hardening of the spectral slope above 10 keV
compared to the 2--10 keV spectrum.  Many radio sources (including Cen
A) appear to have significantly weaker reflection features than do
Seyfert nuclei (Wo\'{z}niak et al.\markcite{wozniak} 1998, Zdziarski et
al.\markcite{aaz} 1995, Grandi et al.\markcite{grandi} 1997).  Indeed,
in most radio-loud objects there is no strong Compton reflection
continuum, and the iron line is interpreted as originating in
Thompson-thin circumnuclear material. In addition, for the case of Cen
A, the spectrum extends to 1 MeV (Steinle et al.\markcite{steinle}
1998), while in radio-quiet Seyferts there is no indication of flux
beyond a few hundred keV (Johnson et al.\markcite{neil} 1997).  This, too,
may indicate a difference in the flow of matter in the vicinity of the
accretion disk, such as advection dominated accretion flow versus thin
disk accretion, and the presence of the collimated radio jet.

Previous observations have shown Cen A to have highly complex
X-ray/gamma-ray properties (D\"{o}bereiner et al.\markcite{dob} 1996;
Turner\markcite{jane} et al.  1997a; Feigelson\markcite{feigelson} et
al. 1981) with multi-temperature diffuse flux, a spatially resolved jet
in radio and X-rays, and a nuclear component with complex low energy
absorption.  The observed iron line at 6.4 keV of equivalent width
$\sim$100 eV is consistent with reprocessing from the absorbing
material (Turner\markcite{jane} et al.  1997a, Sugizaki et
al.\markcite{sugi} 1997).  At energies above $\sim$ 10 keV, Wo\'zniak
et al.\markcite{woz} (1998) analyzed {\it Ginga} data and found no
evidence for a strong Compton reflection component ($\Omega$/2$\pi <
$0.15) in addition to the heavily absorbed
(N$_H$=1.7$\pm0.1\times$10$^{23}$cm$^{-2}$) power law
($\Gamma$=1.79$^{+0.06}_{-0.05}$). This implies there is little cold,
Thompson-thick material in the close vicinity of the AGN (Wo\'{z}niak
et al.\markcite{woz} 1998) that reflects X-rays into our line of
sight.  Taken together with the iron line equivalent width, this result
is consistent with transmission of the primary flux through the
line-of-sight absorber and viewing the accretion disk nearly edge-on.
{\it CGRO}/OSSE observations reveal a steepening of the spectrum at
140--170 keV, which might be due to a nonthermal electron distribution
in the central engine (Kinzer et al.\markcite{kinzer} 1995). The {\it
CGRO}/COMPTEL and {\it CGRO}/EGRET instruments indicate that this
steepening trend continues to 1 GeV with spectral indices ranging from
2.6 to 3.3 depending upon intensity (Steinle et al.\markcite{steinle}
1998).

In this paper we present the results of a short 10 ks observation of
Cen A by the Rossi X-ray Timing Explorer ({\it RXTE}) in the range
2.5$-$240 keV. Section 2 details the observations, section 3 presents
the data reduction and analysis, section 4 discusses the the results
with respect to Cen A itself and in the larger view of AGN unification,
and section 5 presents the conclusions.

\section{Observations}

The {\it RXTE} observed Cen A from 2:43
to 8:05 UT on 1996 August 14 with the Proportional Counter Array (PCA)
and the High Energy X-ray Timing Experiment (HEXTE). The PCA contains 5
large area, xenon, multiwire, multilayer, gas proportional counter
units that cover the energy range 2-60 keV with energy resolution of 1
keV at 6 keV (Jahoda et al.\markcite{jahoda} 1996). All 5 proportional
counters were used in this observation for a total open area of
$\sim$7000 cm$^2$. The field of view of each PCA unit is constrained to
1$^\circ$ Full Width Half Maximum (FWHM) by a beryllium/copper
honeycomb collimator and all 5 are coaligned to within 5 arcminutes to the
spacecraft science pointing axis.

The HEXTE consists of two independent clusters of detectors, each
cluster containing four NaI(Tl)/CsI(Na) phoswich scintillation
counters, each identically collimated by a lead honeycomb (Rothschild
et al.\markcite{rer} 1998).  All eight collimators are co-aligned with the PCA
on-source to give both clusters a 1$^{\circ}$ FWHM field of view.  The
net open area of the eight detectors is $\sim$1600 cm$^2$, and each
detector covers the energy range 15-250 keV with an average energy
resolution of 9 keV at 60 keV.

Throughout the {\it RXTE} mission, the All-Sky Monitor (ASM) provides
an estimate of the 1.5-12 keV flux from the 100 brightest X-ray sources
on a 90 minute and a daily basis.  The ASM is composed of three
scanning shadow cameras, each of which is a position sensitive xenon
proportional counter that views a 6$^\circ \times$90$^\circ$ FWHM
section of sky through a one-dimensional coded mask (Levine et al.\markcite{asm} 
1996). Each camera has $\sim$30 cm$^2$ at 5 keV.

\begin{figure} 
\plotfiddle{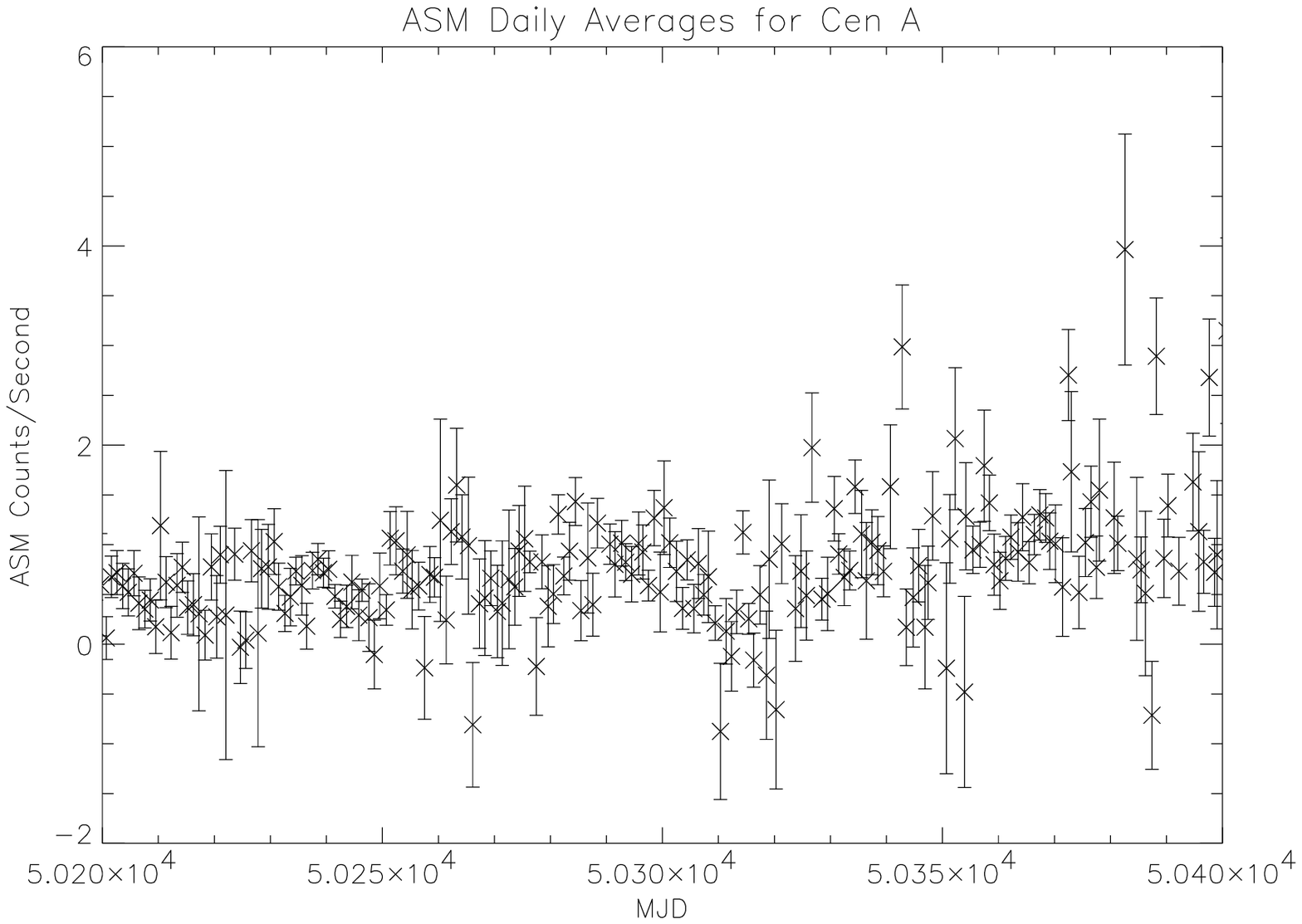}{4in}{0.}{100.}{65.}{-324}{-150}
\plotfiddle{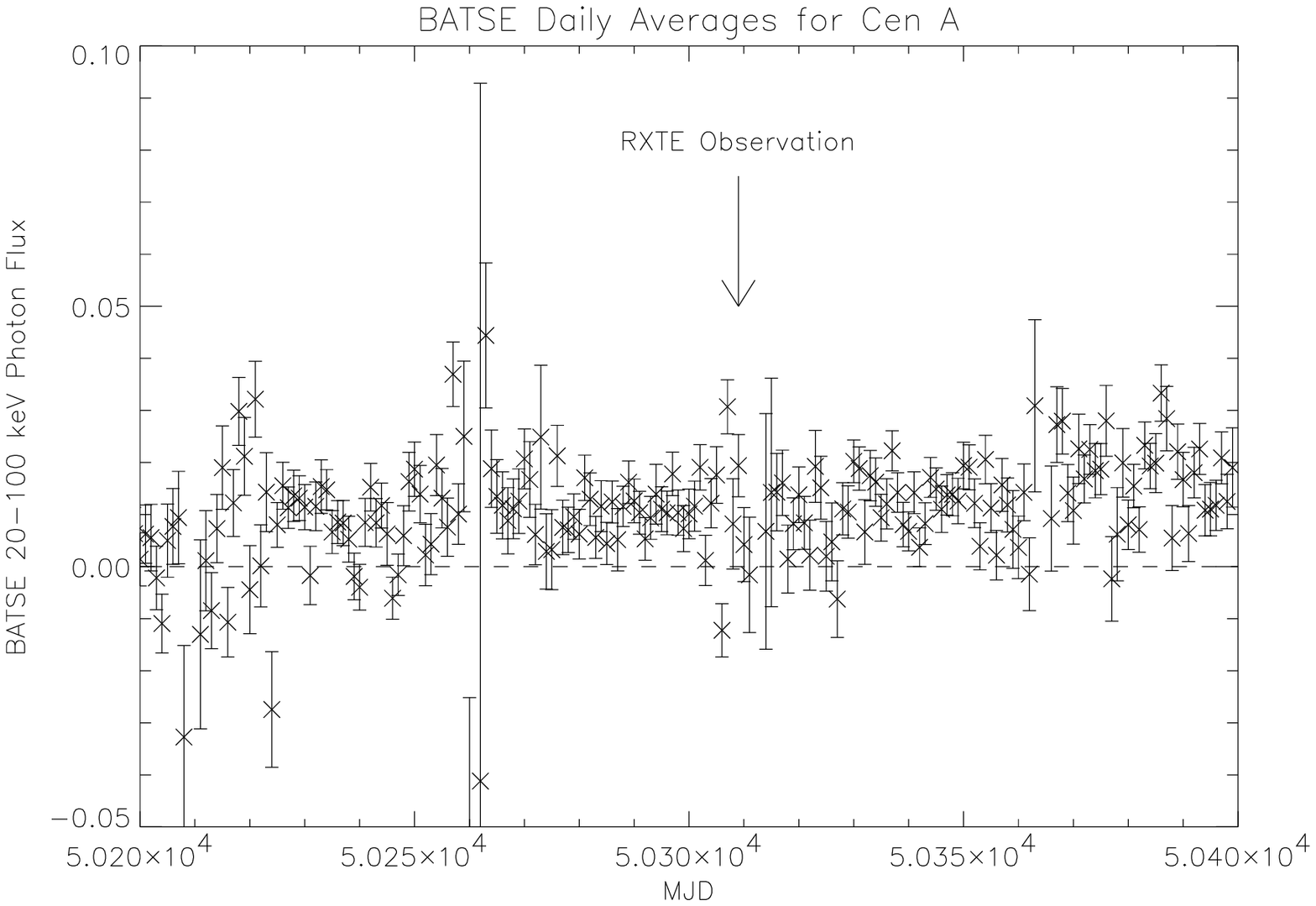}{4in}{0.}{100.}{65.}{-324}{-150}
\caption{{\it RXTE} All-Sky Monitor (ASM) daily average counting
rates in the 2-12 keV band and the {\it CGRO}/BATSE Earth-occultation fluxes
in the 20-100 keV band  for Cen A during 200 days centered on the {\it
RXTE} pointed observation. The arrow indicates the date of the Cen A 
observation.\label{fig.asm}}
\end{figure}

Figure \ref{fig.asm} shows the {\it RXTE} ASM counting rate for Cen A
in the 1.5-12 keV range for 200 days centered on the {\it RXTE} pointed
observation.  The day of the observation (MJD 50309) is indicated in
the Figure, and is at or shortly after the beginning of an interval of
lower than average flux. Comparison of PCA and ASM fluxes indicates the
PCA/HEXTE observations were made at or near the lower flux level. The four
PCA/HEXTE viewing intervals, separated by Earth-occults and South
Atlantic Anomaly (SAA) passages do not reveal any significant change in
intensity over the 5.5 hours of the observation.  Also shown are the
20-100 keV fluxes measured on a daily average by {\it CGRO}/BATSE using
the Earth occultation technique.  The BATSE data are not sensitive to
the magnitude of temporal variation seen by the ASM, as evidenced by
the BATSE error bars.  Thus, the {\it RXTE} observation appears to have
occurred during a temporary period of lower intensity than for the few
weeks on either side of the observation. While this rate is low vis-a-vis
the weeks surrounding the observation, it is quite typical when compared to
past observations.

\section{Data Reduction and Analysis}

The PCA standard data were used for the accumulation of spectra and
light curves, while the HEXTE standard data provided the HEXTE spectral
histograms and light curves. The data from both instruments were
accumulated on 16 s intervals and no significant variability was noted.
The PCA response matrix used was dated 24 April 1998. Estimates of
systematic errors were added to the statistical errors for the PCA
spectral data to reduce the effects of imperfect knowledge of the
instrument response near the K- and L-edges of xenon. These errors
were: 0.5\% from 2.5 -- 5.5 keV, 1\% from 5.5 -- 8 keV, 0.5\% from 8 --
20 keV, and 3\% from 20 -- 90 keV.  These values were determined from
residuals to fits to the Crab nebula observations.  No additional
systematic errors were incorporated in the HEXTE data, because the
statistical errors were larger than the few percent systematic errors
reported in Rothschild et al.\markcite{rothschild} (1998).

\subsection{Background Subtraction}

The PCA background was estimated using models of the instrument
background, activation, and the cosmic diffuse flux derived from blank
sky observations obtained throughout 1997.  We were able to provide a
good background estimate over the entire PCA range by selecting non-SAA
and non-occulted data, and removing time intervals of electron
precipitation as evidenced by raw rates in various layers of the PCA.
The PCA analysis was performed over the 2.5$-$60 keV range. PCA source
and background rates in various energy channels are given in
Table 1.

Contrary to lower energy measurements, hard x-ray observations are made
in a background-dominated regime, except for the brightest sources.
The background, therefore, must be measured during each observation on
timescales short compared to those for background variations. Modeling
background from counting rates is at least an order of magnitude less
accurate, and is not appropriate for hard x-ray measurements.  The
HEXTE cluster rocking subsystem moves the viewing direction of the
detector arrays in a cluster between on-source and off-source every 16
s in order to obtain the near-real time measurement of the instrument
background.  The nominal angular offset of 1.5$^{\circ}$ on either side
of the on-source position was used in the Cen A observation. Since the
rocking axes of the two clusters are orthogonal to each other, four
background regions were sampled around Cen A's position. No confusing
sources were detected. This technique provides background subtractions
that are not based upon any modeling, and are direct measurements taken
during each and every observation.  Analysis of blank fields
demonstrated that the net blank-field flux was consistent with zero
with residuals of 1\% of background or less (Rothschild et
al.\markcite{rer} 1998). The HEXTE analysis was performed over the
15$-$240 keV range. HEXTE source and background rates in various energy
channels are given in Table 1.

\subsection{Spectrum}

\begin{figure}
\plotfiddle{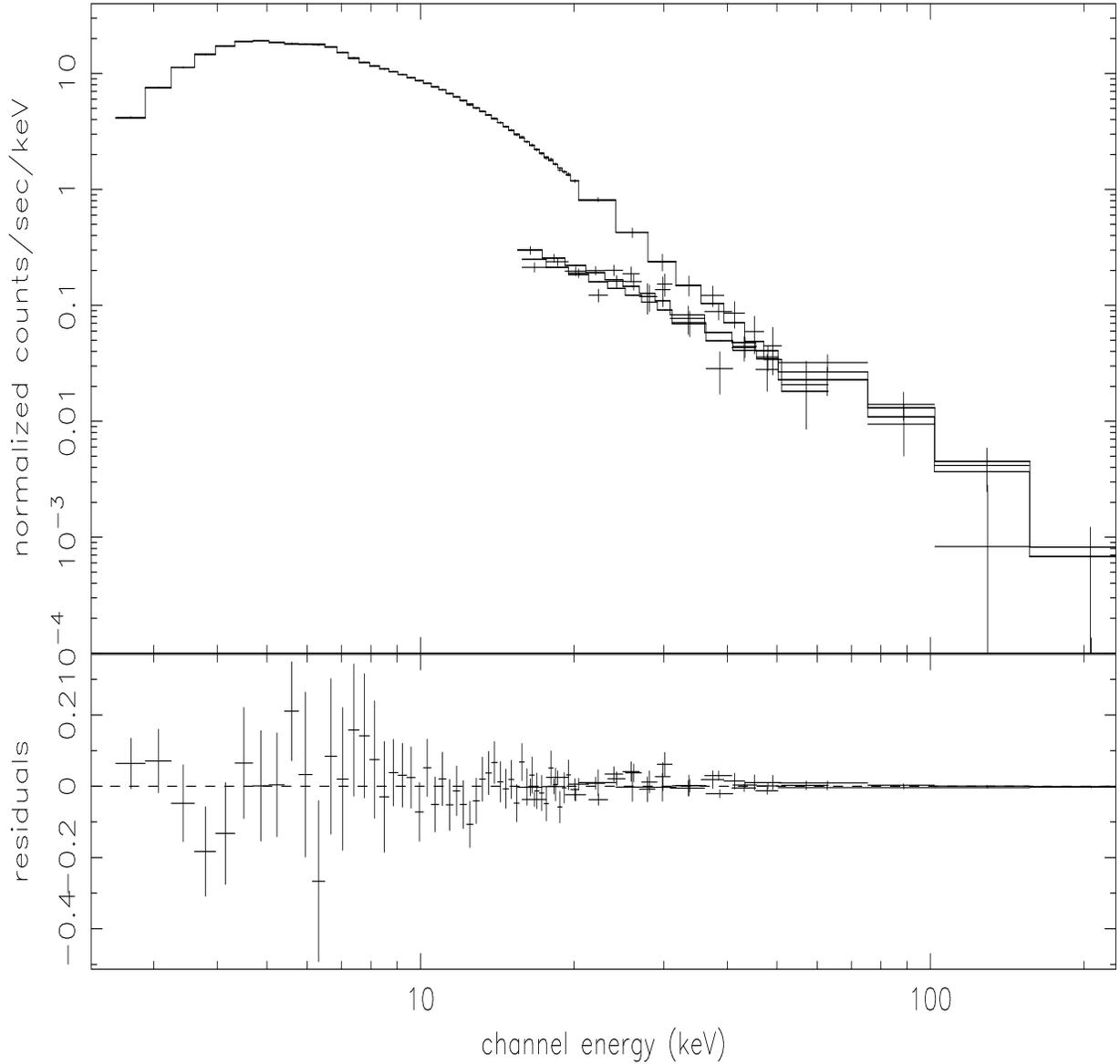}{6in}{270.}{70.}{100.}{-252}{544}
\caption{PCA and HEXTE spectral histograms plus best fit model for
the {\it RXTE} observation of Cen A. The lower panel shows the
residuals to the fit. The relative normalization of each HEXTE cluster
has been left as a free parameter in the fitting process. The
difference between the two HEXTE sets of data are the result of
having data from only 3 of the 4 detectors in cluster B. The HEXTE high 
energy data has been rebinned for display purposes\label{fig.hst}}
\end{figure}

\begin{figure}
\plotfiddle{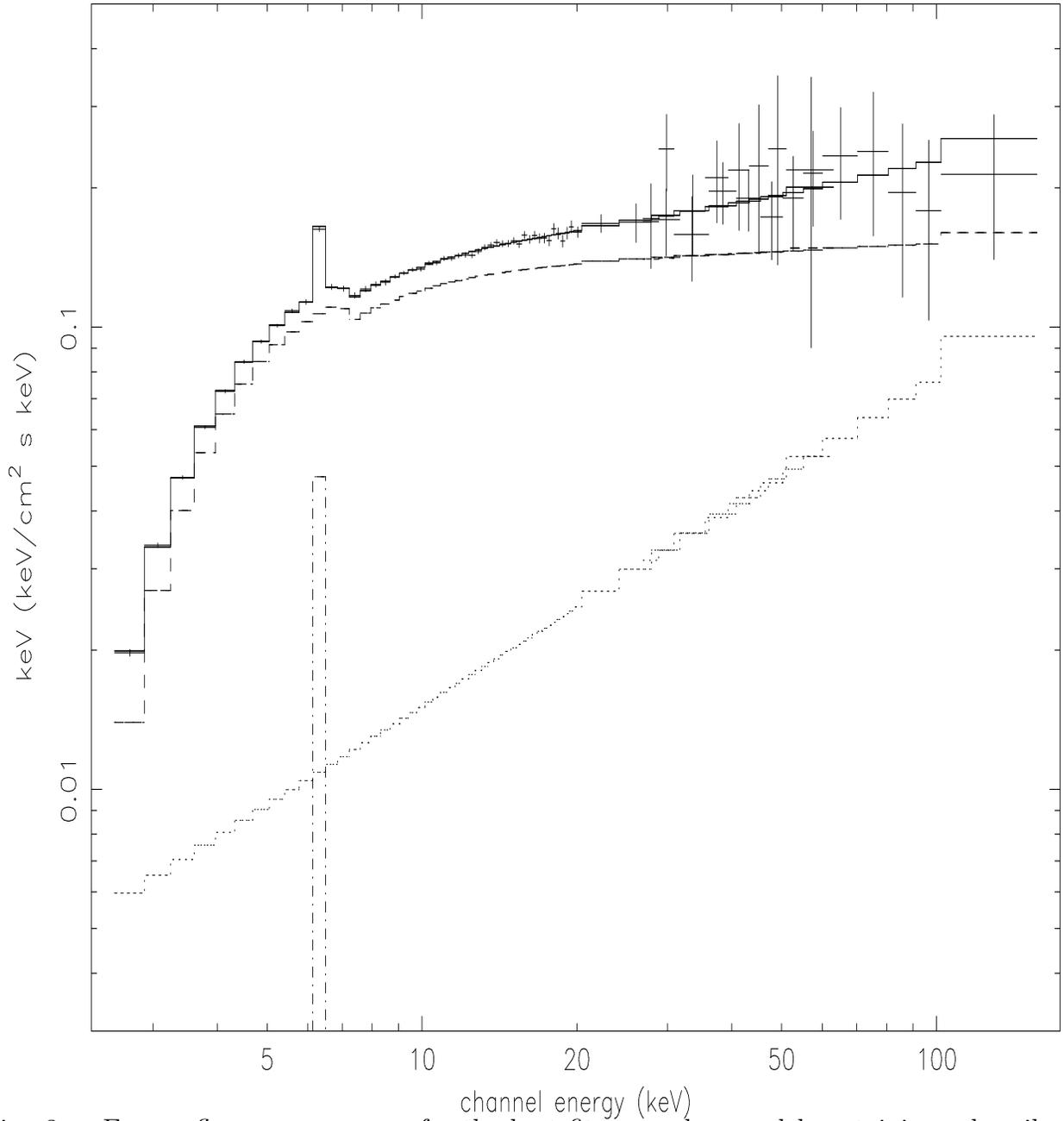}{6in}{270.}{70.}{100.}{-252}{544}
\caption{Energy flux versus energy for the best fit power law
model containing a heavily absorbed power law and a broad iron line,
plus a non-absorbed power law. The HEXTE data have been combined into a
single spectrum and normalized to the PCA flux. Both PCA and HEXTE data
have been rebinned for display purposes.\label{fig.nufnu}}
\end{figure}

The PCA and HEXTE data were fit simultaneously using the XSPEC 10.0
package with the relative normalization of the two instruments included
as a fitted parameter. We utilized a model that contained a heavily
absorbed power law plus gaussian emission line to represent the primary
flux.  The best-fit values for the case of the emission line width set
to zero is given in Table 2. All error bars, unless specified
differently, are 95\% confidence, and generated by varying one
parameter and fitting all others. The decrease in $\chi^2$ that
occurred when the line width was included as a free parameter was 5.2
for one additional degree of freedom out of 540. The F-test indicates
that the probability of this being a fluctuation is 0.021, and
therefore not significant at the 98\% level.  The range of the iron
line width at the 95\% confidence level is 0.11~keV$<
\sigma$(FWHM)$<$0.54~keV.  The energy of the line, 6.46$\pm$0.097 keV,
indicates that the emission is from cool, i.e. weakly ionized,
material. The best fit model with a narrow line is shown in
Figure~\ref{fig.hst} as the predicted instrument counts histogram
compared to the observed counts histogram.  This figure has not been
corrected for the relative normalization of the two HEXTE clusters with
respect to the PCA (see Table 2).  The inferred incident spectrum is
shown in Figure ~\ref{fig.nufnu} in the form of $\nu$F$_\nu$ versus
energy with the HEXTE flux normalized to the PCA and with the points
from both detector systems rebinned for display purposes. The data from
the two HEXTE clusters have also been combined into a single spectrum.

Turner et al.\markcite{jane} (1997a) reported an additional component
due to ROSAT observations of the jet from Cen A, characterized by a
power law with photon index $\Gamma$=2.29$^{+0.67}_{-0.46}$ or a
thermal bremsstrahlung model with kT=1.27$^{+1.21}_{-0.44}$. We tested
the Cen A spectrum for the presence of an additional unabsorbed
component in two ways. The first was by the addition of a power law
whose photon index was a free parameter in the fitting. The decrease in
$\chi^2$ was 4.0 for 2 additional parameters.  The F-test indicates a
probability of 0.13 that a fluctuation could have produced such a
value.  The second test was to utilize a partial covering fraction
model of the absorber that would allow for a percentage of the primary
flux to be transmitted through the absorber.  In this case the $\chi^2$
decreased by 7.4, which, again, is less than a 3$\sigma$ improvement.
The 95\% confidence range is 1.3--6.6\% for the fraction  of the
primary flux that is transmitted. As a result, we conclude that an
additional unabsorbed component was not required by the short {\it
RXTE} observation.

The {\it RXTE} data were tested for the presence of a spectral break, or
steepening at higher energies, using a broken power law. The best-fit
broken power law model was consistent with no break at all, and the
power law index above the break energy (held fixed at 140 keV) was
essentially unconstrained.  This result is due to the small number of
counts in HEXTE above 100 keV in this short observation.  This is
consistent with the {\it CGRO}/OSSE (Kinzer et al.\markcite{kinzer} 1995) and
Welcome-1 (Miyazaki et al.\markcite{miya} 1996) results which extent to
higher energies.

In order to test for the presence of a Compton reflection component,
two fits were made to the data. First, an absorption edge at 7.1 keV
was added to the heavily absorbed primary power law. This resulted in
an upper limit to the absorption depth of $\tau\leq$0.016, indicating
that no strong reflection component was present. The lack of an iron
absorption edge in the data also implies that improper modeling of the
continuum does not affect the inferred width of the fitted Fe line (see
Zdziarski, Johnson, \& Magdziarz\markcite{zjm} 1996 for a discussion of
this systematic effect in NGC 4151). This was further confirmed by
substituting a power law plus Compton component (XSPEC 10.0 model
PEXRAV, Magdziarz \& Zdziarski\markcite{magd} 1995) for the primary
power law in the model.  The 90\% upper limit to a reflection
component, expressed as the ratio of the solid angle for primary flux
scattered into our line of sight to an infinite slab, was found to be
$\Omega$/2$\pi\leq$0.088 for an assumed inclination angle of
70$^\circ$.

When we modeled the iron line as coming from a relativistic accretion
disk with an r$^{-2}$ dependence of the emissivity from 10 GM/c$^2$ to
1000 GM/c$^2$, we found the disk inclination not to be constrained by
the observations (90\% confidence interval includes all angles from 0
to 90 degrees) and the resulting centroid was
consistent with the gaussian line fits. The other model parameters are
not substantially changed, and the $\Delta\chi^2$=5.0 with respect to
the narrow line model was not significant (probability=0.024). Thus,
two indications of observed X-ray emission from an accretion disk -- a
Compton reflection spectrum and an iron line from a disk -- are not
required by the present data.

To recapitulate, the {\it RXTE} spectrum of Cen A can be best described
by a heavily absorbed power law extending above 100 keV and a
162$\pm$24 eV equivalent width iron line from cold matter. The FWHM of
the iron line is less than 0.54 keV, and therefore taken to be narrow.
Measurement of a steepening in the power law above 100 keV is precluded
by the small number of counts in that range for this short
observation.  Neither an additional unabsorbed low energy component,
nor a Compton reflection component are required to fit the data.

The {\it RXTE} best fit can be compared to the {\it ASCA} (Sugizaki et
al.\markcite{sugi} 1997; Turner et al.\markcite{jane} 1997a) and the
{\it Ginga} (Wo\'{z}niak et al.\markcite{woz} 1998) results. The {\it
RXTE} power law index of 1.857$\pm$0.008 agrees with the {\it ASCA} and
{\it Ginga} values. The measured absorption, however, has decreased
about 60\% from $\sim$1.6$\times$10$^{23}$ cm$^{-2}$ to the {\it RXTE}
value of 0.94$\pm$0.01$\times$10$^{23}$ cm$^{-2}$ in three years. At
the same time the equivalent width of the narrow iron line has
increased 33\% from $\sim$120 eV to $\sim$160 eV.  The flux of the Fe
line, on the other hand, has not changed significantly (see Table 3).
The {\it RXTE} and {\it Ginga} upper limits on the reflection component
are in complete agreement, i.e., none is detected to less than
$\Omega$/2$\pi \sim$0.09. This can be compared to a weighted mean of
0.72$\pm$0.05 for 27 {\it GINGA} Seyfert galaxies (Nandra \&
Pounds\markcite{nandra} 1994).  The latter sample was dominated by
radio-quiet objects with only three 3C sources included.

\section{Discussion}

Observations of molecular lines in emission and absorption from Cen A
(Israel et al.\markcite{israel} 1990) imply that a thick torus with
outer radius of 325 pc, inner radius of 80 pc and thickness of 80 pc
surrounds the nucleus. The density of the torus decreases as r$^{-2}$
from an inner edge value of 2$\times$10$^4$ cm$^{-3}$.  With an
estimated mass of 2$\times$10$^7$ M$_\odot$, the line of sight through
the torus would have $\sim$1.0$\times$10$^{23}$ atoms cm$^{-2}$, which
is consistent with past and present measurements in the X-ray range.
Leahy \& Creighton\markcite{leahy} (1993) have simulated X-ray spectra
for power law fluxes embedded in a spherical absorber, and Figure 1 of
Turner et al.\markcite{janeb} (1997b) displays the result for a power
law photon index of 2.  This calculation predicts about 100 eV
equivalent width for a column density of $\sim$1$\times$ 10$^{23}$
cm$^{-2}$.  If the iron line were due to reprocessing of the primary
flux in a cold accretion disk by Compton reflection, on the other hand,
equivalent widths close to 300 eV would be expected.  The present Cen A
result of 162$\pm$24 eV equivalent width and no evidence of a Compton
component is consistent with the iron emission coming from a power law
flux embedded in a uniform spherical absorber.

Over the last two decades, the flux from the iron line and the continuum
flux have been measured by instruments on {\it OSO-8}, {\it EXOSAT},
{\it Tenma}, {\it Ginga}, {\it ASCA}, and now {\it RXTE}. Table 3 gives
the published history of these measurements of the spectral index,
absorbing column, iron line equivalent width, iron line line flux, and
2-10 keV flux from Cen A.  The multiple observations by {\it OSO-8} and
{\it EXOSAT} each claim no variation in iron line flux, while reporting
changes in 2$-$10 keV continuum flux of factors of 2 or more. On the
other hand, both the line and continuum fluxes are reported to be
essentially constant for the two {\it Ginga} observations. Over the
last 20 years the absorbing column depth has ranged over 1--2 $\times$
10$^{23}$ cm$^{-2}$ while the equivalent width has values from $<$50 to
$>$200 eV. Comparison to the Leahy and Creighton\markcite{leahy} (1993)
model displayed in Figure 1 of Turner et al.\markcite{janeb} (1997b),
shows that the equivalent width/absorbing column values scatter about
the calculation for circumnuclear material with, at most, factor of two
deviations above and below the model predictions.

Does the iron line flux vary? Fitting the data in Table 3 to a single
mean value of 4.65$\times$10$^{-4}$ photons cm$^{-2}$ s$^{-1}$ yields a
$\chi ^2 _\nu$ = 1.15 for 10 degrees of freedom. This implies the line
fluxes are consistant with a single value. At the same time, the mean
continuum flux is 3.86$\times$10$^{-10}$ ergs cm$^{-2}$ s$^{-1}$ with a
large scatter about this value. Fitting to this mean value yields a
$\chi ^2 _\nu$  in the thousands.  Clearly, the iron line originates
far away from the nucleus.  Due to the sparse sampling of Cen A by
instruments capable of measuring the iron line and continuum flux, one
cannot test for time delays between continuum variations and those in
the line flux, which might support the molecular torus as the site of
the iron line emission.

Using {\it ASCA} data, Sugizaki et al.\markcite{sugi} (1997) found a hard
nuclear and a soft diffuse component plus low-Z lines of Mg, Si, and S
and the high-Z line from Fe. The soft component had a factor of ten
less absorbing column depth than the hard component, indicating that it
was not being viewed through the torus.  The equivalent widths of the
low-Z lines and the soft continuum were compatible with being produced
by the primary hard flux scattered by cold matter. The iron line
equivalent width, under the same assumptions, was a factor of two
larger than expected. From this, Sugizaki et al.\markcite{sugi} (1997)
concluded that the low-Z emitting material and the material responsible
for the Fe line must be located in separate regions.  This further
implies that the covering factor of the primary source must be less
than unity for the primary flux to reach the reprocessing site.  This
is consistent with the unified model where the molecular torus does not
cover the poles but does impact observations of the primary flux from
high inclination angles.

Nandra et al.\markcite{nandraetal} (1997) have studied 18 Seyfert 1
galaxies with {\it ASCA} and find an average spectral behavior, using the
relativistic disk model, that is compatible with unification models,
i.e., photon indices averaging 1.9, equivalent widths of iron lines
averaging 230 eV, and small ($\sim$30$^\circ$) inclination angles.  The
line widths are explained by emission from a range of radii in the
accretion disk around a supermassive black hole, and emission from
another region is not required. Importantly, another component to the
iron emission due to observing the primary flux through a dense (N$_H
\approx$ 10$^{23}$ atoms cm$^{-2}$) uniform medium is not required.
Nandra and Pounds\markcite{nandra} (1994) used the broad-band {\it Ginga}
observations of 20 Seyfert 1 and 7 Narrow Emission Line Galaxies, that
are also classified as Seyfert 2 galaxies (Turner et
al.\markcite{janeb}, 1997b), to investigate reflection components under
the assumption of a face-on geometry. Reflection components were detected in
roughly half of the samples of both types of Seyfert galaxy.

Turner et al.\markcite{janeb,janec} (1997b,c) have summarized {\it
ASCA} observations of 25 Seyfert 2 galaxies and thoroughly discussed
the consequences of observing scattered and reflected X-rays. Several
radio quiet galaxies in their sample exhibited iron lines (EW$>$200 eV)
from cold (E$_{Fe}$=6.4 keV) material, and yet had little or no
evidence for Compton reflection components, which is consistent with
viewing the accretion disk at a high inclination angle. When combined
with the Seyfert 1 observations, this indicates that the inner regions
of near edge-on radio galaxies and radio quiet Seyfert 2 galaxies are
similar, and that the unified model of active galaxies may be
applicable independent of radio character.  This further implies that
the presence of the highly collimated jet in radio galaxies like Cen A
plays a small part in determining the disk/reflection emission from
active galaxies.  On the other hand, when one is viewing within the
collimated beam, as is the case for blazars, the jet dominates the
x-ray and gamma-ray emission observed.

\section{Conclusions}

A short {\it RXTE} observation of Cen A during a short period of
reduced flux has revealed no temporal variability over 5.5 hours and a
complex spectrum in the 2-250 keV range. The best fit spectral model
consists of a heavily absorbed primary flux and an emission line from
cool iron.  The measured column depth, iron line equivalent width, and
power law index are consistent with earlier observations. The iron line flux
has not been observed to vary in sparce sampling of it over the last 
two decades, while the continuum flux has varied dramatically, indicating
that the iron line emission occurs far from the primary X-ray emission site.

The line equivalent width is consistent with production in the
molecular torus by the primary hard X-ray flux, and the weakness or
lack of a Compton reprocessing component is consistent with the nearly
edge-on view of the accretion disk.  In addition, the lack of a line at
6.7 keV and no Compton component indicates that an ionized gas
enveloping the broad line region is not scattering a significant amount
of X-ray flux into our line of sight. Cen A's X-ray character is
essentially the same as that of radio-quiet Seyfert 2 galaxies. This
indicates that the presence of the collimated radio jet does not
significantly affect the parameters of the matter within the inner
parsec of the system and that the unification model for radio-quiet AGN
may be extended to include low luminosity radio galaxies. Future
observations of Cen A will be made in 1998 and will be the basis for
variability studies that may shed more light on the geometry of the
inner region of this active galaxy.

\acknowledgments

We acknowledge the excellent work of the {\it RXTE} Science Operations
Center staff to provide the observations and the Guest Observer
Facility for providing support in analyzing them.  We also thank Dr.
Alan Harmon and the BATSE instrument team for providing the BATSE data
on Cen A, and the thoughtful comments from the referee. This work was 
supported by NASA contract NAS5-30720.

\clearpage
\begin{deluxetable}{rrrr}
\footnotesize
\tablecaption{PCA and HEXTE Counting Rates}
\tablewidth{0pt}
\startdata
\underline{PCA (5 detectors)} & \underline{Cen A} & \underline{Background} & \underline{Cen A/Background}\nl
2--10 keV & 100.5$\pm$0.25 c/s & 31.45$\pm$0.05 c/s & 3.20$\pm$0.008\nl
10--20 keV & 37.71$\pm$0.11 c/s & 25.27$\pm$0.05 c/s & 1.49$\pm$0.004\nl
20--60 keV & 7.25$\pm$0.35 c/s & 67.57$\pm$0.08 c/s & 0.11$\pm$0.005\nl
&\nl
\underline{HEXTE A (4 detectors)}\nl
15--30 keV & 2.99$\pm$0.13 c/s & 24.57$\pm$0.10 c/s & 0.122$\pm$0.005\nl
30--100 keV & 2.38$\pm$0.24 c/s & 80.23$\pm$0.18 c/s & 0.030$\pm$0.003\nl
100--240 keV & 0.24$\pm$0.20 c/s & 54.88$\pm$0.15 c/s & 0.004$\pm$0.004\nl
&\nl
\underline{HEXTE B (3 detectors)}\nl
15--30 keV & 2.21$\pm$0.11 c/s & 15.76$\pm$0.08 c/s & 0.140$\pm$0.007\nl
30-100 keV & 1.98$\pm$0.21 c/s & 60.14$\pm$0.16 c/s & 0.033$\pm$0.003\nl
100--240 keV & 0.04$\pm$0.17 c/s & 38.86$\pm$0.13 c/s & 0.001$\pm$0.004\nl
\enddata
\end{deluxetable}

\clearpage
\begin{deluxetable}{lll}
\footnotesize
\tablecaption{Fit to PCA (2.5--60 keV)/HEXTE (15--240 keV) Spectra}
\tablewidth{0pt}
%\tablehead{& \colhead{Broad} & \colhead{Narrow} & & \\
%\colhead{Parameter} & \colhead{\underline{Gaussian Line}} & \colhead{\underline{Gaussian Line}} & \colhead{\underline{Disk Line}} & %\colhead{\underline{Reflection}}}
\startdata
& --- Best Fit Spectrum --- \nl
\hline
Absorption & N$_H$ = 9.42$\pm$0.24 $\times$10$^{22}$ cm$^{-2}$\nl
Power Law & $\Gamma$ = 1.857$\pm$0.015 \nl
          & F(2--10 keV) = 3.398$\pm$0.052 $\times$10$^{-10}$ ergs cm$^{-2}$ s$^{-1}$ \nl
Iron Line & E$_{Fe}$ = 6.46$\pm$0.097 keV \nl
          & $\sigma$ (FWHM) = 0. \nl
          & F(6.46 keV) = 5.37$\pm0.83$ $\times$10$^{-4}$ photons cm$^{-2}$ s$^{-1}$ \nl
          & E.W. = 162$\pm$25 eV \nl
Normalization & cluster A\tablenotemark{a} = 0.688$\pm$0.020 \nl
              & cluster B\tablenotemark{b} = 0.693$\pm$0.024 \nl
              & Flux(2--10 keV) = 1.972$\pm$0.002 $\times$10$^{-1}$ ergs cm$^{-2}$ s$^{-1}$ \nl
Fit       & $\chi^2$/D.O.F. = 517.2/541 \nl
          & $\chi^2_\nu$ = 0.956 \nl
\enddata
\tablenotetext{a}{Ratio of HEXTE to PCA Normalization for Cluster A}
\tablenotetext{b}{Ratio of HEXTE to PCA Normalization for Cluster B}
\end{deluxetable}

\begin{deluxetable}{llllllll}
\footnotesize
\tablecaption{Previous Observations of the Iron Line in Cen A}
\tablewidth{0pt}
\tablehead{
\colhead{Satellite} & \colhead{Observation Date}& \colhead{Photon Index}& 
\colhead{Absorption\tablenotemark{a}}& \colhead{E.W.\tablenotemark{b}} & \colhead{Line Flux\tablenotemark{d}} &  Source Flux\tablenotemark{f} & \colhead{Ref}}
\startdata
OSO-8  & Jul 27$-$Aug 5, 1975& 1.66$\pm$0.03\tablenotemark{c} & 1.33$\pm$0.07\tablenotemark{c} & 120$\pm$35 \tablenotemark{c} & 5.7$\pm$1.7\tablenotemark{e} & 4.66$\pm$0.06& 1\\
OSO-8  & Jul 28$-$Aug 8, 1976& 1.66$\pm$0.03\tablenotemark{c} & 1.33$\pm$0.07\tablenotemark{c} & 120$\pm$35
\tablenotemark{c} & 5.7$\pm$1.7\tablenotemark{e} & 2.39$\pm$0.11 & 1\\
EXOSAT & Feb 13, 1984         & 1.75$\pm$0.08 & 1.64$\pm$0.08 & 164$\pm$36 & 4.7$\pm$1.0 & 3.4$\pm$0.5 & 2\\
TENMA  & Mar 29$-$Apr 4, 1984 & 1.80$\pm$0.04 & 1.53$\pm$0.03 &  90$\pm$20 & 5.5$\pm$1.2\tablenotemark{e} &7.6$\pm$0.07 & 3\\
EXOSAT & Jun 8, 1984         & 1.57$\pm$0.10 & 1.40$\pm$0.10 & 207$\pm$52 & 3.7$\pm$0.9 & 2.6$\pm$0.5 & 2\\
EXOSAT & Jul 30, 1984        & 1.61$\pm$0.08 & 1.70$\pm$0.09 & 145$\pm$43 & 3.1$\pm$0.9 & 2.9$\pm$0.4 & 2\\
EXOSAT & Jun 29, 1985        & 1.70$\pm$0.02 & 1.44$\pm$0.02 &  63$\pm$9  & 5.7$\pm$0.8 & 11.3$\pm$0.3 & 2\\
GINGA  & Mar 8$-$9, 1989      & 1.80$\pm$0.02 & 1.61$^{+0.07}_{-0.05}$ & 110$\pm$32 & 5.2$^{+0.9}_{-1.0}$ & 2.3$\pm$0.1 & 4\\
GINGA  & Feb 7$-$9, 1990       & 1.81$\pm$0.02 & 1.43$\pm$0.03 & 110$\pm$32 & 5.2$\pm$0.9 & 2.3$\pm$0.1 & 4\\
ASCA   & Aug 14, 1993         & 1.83$^{+0.5}_{-0.4}$ & 1.29$\pm$0.40 & 68$\pm$39 & 3.07$\pm$1.72\tablenotemark{e} & 1.78$\pm$0.18 & 5\\
ASCA   & Aug 14$-$15, 1993    & 1.96$\pm$0.01 & 1.10$^{+0.02}_{-0.01}$ (59\%) & 114$\pm$18 & 3.91$^{+0.46}_{-0.47}$ & 1.8$\pm$0.06 & 6\\
&&& 3.54$^{+0.20}_{-0.24}$ (40\%)\\
RXTE   & Aug 14, 1996         & 1.86$\pm$0.015 & 0.94$\pm$0.02 & 162$\pm$25 & 5.37$\pm$0.83 & 3.40$\pm$0.05 & 7\\
\enddata
\tablenotetext{a}{$\times$10$^{23}$ cm$^{-2}$} 
\tablenotetext{b}{Equivalent Width in eV}
\tablenotetext{c}{Average of the 1975 and 1976 data}
\tablenotetext{d}{$\times$10$^{-4}$ photons cm$^{-2}$ s$^{-1}$}
\tablenotetext{e}{Calculated from parameters given in reference}
\tablenotetext{f}{$\times$10$^{-10}$ ergs cm$^{-2}$ s$^{-1}$ in the 2$-$10 keV band}
\tablenotetext{1}{Mushotzky et al.\markcite{mushy} 1978}
\tablenotetext{2}{Morini, Anselmo, \& Molteni\markcite{mam} 1989}
\tablenotetext{3}{Wang, B. et al.\markcite{wang} 1986.}
\tablenotetext{4}{Miyazaki et al.\markcite{miya} 1996}
\tablenotetext{5}{Sugizaki, M. et al.\markcite{sugi} 1997}
\tablenotetext{6}{Turner et al.\markcite{jane} 1997a}
\tablenotetext{7}{present work}
\end{deluxetable}

\clearpage

\clearpage

\end{document}